\documentclass[twocolumn,showpacs,preprintnumbers,amsmath,amssymb]{revtex4-1}
\usepackage{graphicx}
\usepackage{rotating}
\begin{document}

\renewcommand{\theequation}{\thesection.\arabic{equation}}

\newcommand{\re}{\mathop{\mathrm{Re}}}

\newcommand{\be}{\begin{equation}}
\newcommand{\ee}{\end{equation}}
\newcommand{\bea}{\begin{eqnarray}}
\newcommand{\eea}{\end{eqnarray}}

%\maketitle

\title{Redshift drift in a pressure-gradient cosmology}

\author{Adam Balcerzak}
\email{abalcerz@wmf.univ.szczecin.pl}
\affiliation{\it Institute of Physics, University of Szczecin, Wielkopolska 15, 70-451 Szczecin, Poland}
\affiliation{\it Copernicus Center for Interdisciplinary Studies, S{\l }awkowska 17, 31-016 Krak\'ow, Poland}

\author{Mariusz P. D\c{a}browski}
\email{mpdabfz@wmf.univ.szczecin.pl}
\affiliation{\it Institute of Physics, University of Szczecin, Wielkopolska 15, 70-451 Szczecin, Poland}
\affiliation{\it Copernicus Center for Interdisciplinary Studies,
S{\l }awkowska 17, 31-016 Krak\'ow, Poland}

\date{\today}

\input epsf

\begin{abstract}
We derive a redshift drift formula for the spherically symmetric inhomogeneous pressure Stephani universes which are complementary to the spherically symmetric inhomogeneous density Lema\^itre-Tolman-Bondi models. We show that there is a clear difference between redshift drift predictions for these two models as well as between the Stephani models and the standard $\Lambda$CDM Friedmann models. The Stephani models have positive drift values at small redshift and behave qualitatively (but not quantitatively) as the $\Lambda$CDM models at large redshift, while the drift for LTB models is always negative. This prediction may perhaps be tested in future telescopes such as European Extremely Large Telescope (EELT), Thirty Meter Telescope (TMT), Giant Magellan Telescope (GMT), and especially, in gravitational wave interferometers DECi-Hertz Interferometer Gravitational Wave Observatory and Big Bang Observer (DECIGO/BBO), which aim at low redshift.

\end{abstract}


\maketitle

\section{Introduction}
\label{intro}
\setcounter{equation}{0}

In the context of the dark energy problem there has been more interest in the non-Friedmannian models of the Universe which could explain the acceleration only due to inhomogeneity \cite{UCETolman,center}. One of the strongest claims was that we were living in a spherically symmetric void of density described by the Lema\^itre-Tolman-Bondi (LTB) dust spheres model \cite{LTB}. However, there are a variety of inhomogeneous models (for a review see e.g. \cite{BKC}) which have the advantage that they are exact solutions of the Einstein field equations and not the perturbations of the isotropic and homogeneous Friedmann models. One of the reasons to investigate the simplest Friedmann models is their mathematical feasibility supported by the paradigm of the Copernican principle which says that we do not live at the center of the Universe. However, observations are practically made from just one point in the Universe (``here and now'') and extend only onto the unique past light cone of the observer on the Earth. It is clear that even the cosmic microwave background (CMB) is observed from such a point. Apparently, its observations prove isotropy of the Universe (isotropy with respect to observation point - the center of symmetry) - but not necessarily homogeneity (isotropy with respect to any point in the Universe) \cite{chrisroy}. Then, the question is whether we should first start with model-independent observations of the past light cone and then make conclusions related to modeling of the Universe (cf. observational cosmology program of Ref.\cite{ellis}). In other words, homogeneity needs a check. Suppose that we have an inhomogeneous model of the Universe with the same number of parameters as a homogeneous dark energy model and they both fit observations very well. How could we differentiate between these two models?

The simplest inhomogeneous cosmological models are the spherically symmetric ones and these are the complementary to each other: the inhomogeneous density $\varrho(t,r)$ (dust shells) LTB models and the inhomogeneous pressure $p(t,r)$ (gradient of pressure shells) Stephani models. Apparently, due to a conservative approach related to the matter content (dust) most of the cosmologists investigate the former, and only a few investigate the latter. In view of large expansion of investigations related to LTB models as nearly the only example of an inhomogeneous cosmology, we think that it is useful to present some geometrical and physical properties of the complementary Stephani models. There are just a few papers about these models in comparison to what has been written about LTB. That is why in this paper we would like to investigate such a complement of LTB models. One of the benefits of Stephani cosmology is that it possesses a totally spacetime inhomogeneous generalization \cite{stephani,dabrowski93} which does not violate the cosmological principle. In fact, we consider our investigations as the first step towards developing more models of such a type - i.e. the universes which describe real inhomogeneity of space (for a review see e.g. Ref. \cite{krasinski}) - not only those which possess a rather unrealistic center of the Universe which is against the Copernican principle. Actually, the Stephani universes were the first inhomogeneous models ever compared with observational data from supernovae \cite{dabrowski98} - and proved that they could be fitted to it. Despite the LTB models being theoretically explored much earlier, only later were they tested observationally \cite{LTBtests}.

Our paper is organized as follows. In Sec. \ref{models} we present some basic properties of inhomogeneous pressure Stephani models, also in comparison to complementary LTB models. In Sec. \ref{exact} we discuss some exact Stephani models useful for further discussion. Section \ref{drift} contains the main result which is the redshift drift in these pressure-gradient cosmologies. In Sec. \ref{conclusion} we give conclusions.

\section{Inhomogeneous pressure Stephani universe}
\setcounter{equation}{0}
\label{models}

The inhomogeneous pressure Stephani model is the only spherically symmetric solution of Einstein equations for a perfect-fluid energy-momentum tensor $T^{ab} = (\varrho + p) u^a u^b + p g^{ab}$ ($p$ is the pressure, $g^{ab}$ is the metric tensor) which is conformally flat (Weyl tensor vanishes) and embeddable in a five-dimensional flat space \cite{stephani,dabrowski93}. A general model has no spacetime symmetries at all, but its three-dimensional hyperspaces of constant time are maximally symmetric like in the Friedmann universe. In order to be consistent with an LTB, here we consider only a spherically symmetric subcase of the Stephani model which reads as (one uses a Friedmann-like time coordinate \cite{dabrowski93})
\bea
\label{STMET}
ds^2~=~-~\frac{a^2}{\dot{a}^2} \left[ \frac{ \left( \frac{V}{a} \right)^{\centerdot}}
  { \left( \frac{V}{a} \right)} \right]^2
%  \frac{a^2}{V^2}
%\left[\left( \frac{V}{a} \right)^{\centerdot} \right]^2
dt^2~
%\nonumber \\&+&~
+ \frac{a^2}{V^2} \left(dr^2~+~r^2 d\Omega^2
%\left(d\theta^2~+~ \sin^2{\theta}d\varphi^2 \right)
 \right),\nonumber \\
 &&
\eea
where
\be
\label{VSS}
  V(t,r)  =  1 + \frac{1}{4}k(t)r^2~,
\ee
and $(\ldots)^{\cdot}~\equiv~\partial/\partial t$. The function $a(t)$
plays the role of a generalized scale factor, $k(t)$ has the meaning of a
time-dependent ``curvature index,'' and $r$ is the radial coordinate. Kinematically, these models are characterized by the nonvanishing expansion scalar $\Theta$ and  the acceleration vector $\dot{u}_a$.

The energy density and pressure are given, respectively, by
\begin{eqnarray}
\label{rhost}
\varrho(t) & = & \frac{3}{8 \pi G} \left[ \frac{\dot{a}^2(t)}{a^2(t)} + \frac{k(t)}{a^2(t)}
\right]~,\\
\label{pst}
p(t,r) & = & \left[ -1 + \frac{1}{3} \frac{\dot{\varrho}(t)}{\varrho(t)} \frac{ \left[ \frac{V(t,r)}{a(t)} \right]}
  { \left[ \frac{V(t,r)}{a(t)} \right]^{\centerdot}} \right] \varrho(t) \equiv w_{e}(t,r) \varrho(t), \nonumber \\
  &&
\end{eqnarray}
where we have set the velocity of light $c=1$, $G$ is the gravitational constant, and $w_{e}(t,r)$ is an effective spatially dependent barotropic index. It is useful that the metric (\ref{STMET}) is written in terms of the isotropic coordinate which is related to a standard Friedmann coordinate $\bar{r}$ via the transformation $\bar{r} = r/(1 + (1/4)kr^2)$ [or its inverse $r = 2\bar{r}/(1 + \sqrt{1-k\bar{r}^2})$]. This allows one to make a good appeal to a general Stephani model with no symmetry at all, which is formulated in Cartesian coordinates \cite{stephani}. Because of having a time-dependent curvature index $k(t)$ in Stephani models, the transformation to the isotropic coordinate is not so convenient as for Friedmann models. The global topology of Stephani models is $S^3 \times R$ and they look like the de Sitter hyperboloid with specific deformations near the ``neck circle'' which is the smallest radius circle of the hyperboloid while taking positive curvature spatial sections \cite{dabrowski93}. The local topology of the constant time hypersurfaces (index $k(t)$) may change in time. In a standard de Sitter case one cuts the hyperboloid by either $k=1$ ($S^3$ topology), $k=0$ ($R^3$ topology) or $k=-1$ ($H^3$ topology). Here we have a ``three-in-one'' case and the Universe may either ``open up'' to become negatively curved or ``close down'' to become positively curved.  In a general Stephani model which has no spacetime symmetry at all, the point reflecting an instantaneous (``only one hypersurface'') center of symmetry moves around a deformed hyperboloid. Similarly as in an LTB model, there are two antipodal centers of symmetry. In the Stephani models there exist instantaneous standard big-bang singularities ($a \to 0$, $\varrho \to \infty$, $p \to \infty$) as well as finite density (FD) singularities of pressure which appear at some particular value of the radial coordinate $r$ \cite{sussmann,dabrowski93}. FD singularities at first glance resemble sudden future singularities (SFS) \cite{barrow04} which appear in Friedmann models with no equation of state to link the energy density and pressure. However, FD singularities occur as singularities in spatial coordinates rather than in time, which means that even at the present moment of the evolution they may exist somewhere in the Universe. It has been shown \cite{PRD05} that SFS may appear in inhomogeneous Stephani universes, independently of the FD singularities. In Stephani models there is also the spacelike $\Pi$ boundary \cite{sussmann} which divides each negative curvature $k(t) < 0$ hypersurface onto the two sheets (the ``far sheet'' and the ``near sheet'' \cite{stephani}). It appears whenever the function $V(t,r)$ in (\ref{VSS}) is zero. On a $\Pi$ boundary the Universe behaves asymptotically like de Sitter. As one can conclude from (\ref{rhost}) and (\ref{pst}), there is no global equation of state; rather it changes from shell to shell, where it is explicit and fixed.

For the sake of comparison we remind that the simplest Lema\^itre-Tolman-Bondi universe is the only spherically symmetric solution of Einstein equations for pressureless matter energy-momentum tensor $T^{ab} = \varrho u^a u^b$ ($\varrho$ is the energy density, $u^a$ is the 4-velocity vector) and no cosmological constant which has a spatially dependent "curvature index" $k(r)$. Models with a nonzero $\Lambda$ term are also possible and can be solved in terms of elliptic functions in analogy to Friedmann models, though with the spatially dependent constants of integration \cite{LTBL}. It has to obey some ``regularity conditions'' like the existence of a regular center of symmetry and the orthogonality of hypersurfaces of constant time (of topology $S^3$ which implies the existence of a second center of symmetry) to a 4-velocity vector (see e.g. \cite{BKC}). Another condition, which in physical terms means the avoidance of the infinite ``spikes'' of density is related to an apparent possibility for ``shell-crossing'' singularities to exist \cite{shell}. However, these singularities are of a weak type in the sense of Tipler and Kr\'olak \cite{tipler+krolak} and are like some recently investigated exotic singularities known as generalized sudden future singularities \cite{GSFS} (of first and higher pressure derivatives or second and higher mass density derivatives) in the Friedmann universes with no geodesic incompleteness \cite{lazkoz} (for a classification of weak singularities in Friedmann cosmology see e.g. \cite{AIP10}).  Kinematically, LTB models are characterized by the nonvanishing expansion scalar $\Theta$ and shear tensor $\sigma_{ab}$. One of the peculiarities is that in LTB models the big-bang singularity is not necessarily instantaneous - different points start evolution in different moments of time.

It should be admitted that the pressureless dust matter present in LTB models has its firm observational basis. However, the acceleration of the Universe forces cosmologists to look for some exotic kinds of matter sources (dark energy). As we know, this is even the case for the cosmological constant, since its observed value is much less than its most common physical interpretation as the vacuum energy. While dealing with Stephani models, we assume an unknown fluid with varying from hypersurface to hypersurface equation of state (which is, however, explicit and fixed on each hypersurface). There have been many proposals for the dark energy and we still do not know what it is. Then, our proposal is on the same footing as many others with possible interpretation as a nongravitational force in the Universe or a kind of a spatially varying cosmological constant (or spatially dependent vacuum energy).

In fact, in Ref. \cite{kam04} Tolman-Oppenheimer-Volkoff (TOV) equilibrium equation for exotic stars made of phantom matter \cite{phantom}
or SFS-related matter, such as that related to a big-brake singularity ($a=$ const., $\dot{a} = 0$ and $\varrho \to 0$, $p \to \infty$) was obtained for Friedmann models  filled with the anti-Chaplygin gas fulfilling the equation of state $p = A^2/\varrho$ ($A=$ const.). In yet another Ref. \cite{kam08} it was found that the Chaplygin gas with $p = - A^2/\varrho$ may serve as a source for stable exotic star configurations which fulfill appropriate TOV equations. Further, Ref. \cite{kam09} shows that the generalized Chaplygin gas model with $p = - A^2/\varrho^{\alpha}$ and a local spherical collapse is a kind of a generalization of the LTB model. It was found that there existed a static spherically symmetric configuration in which the central pressure at $r=0$ was constant, while on some shell of constant radius $r_s$ it became minus infinity. This is an analogue of the FD singularity of Stephani models, though without time evolution. Besides, on an arbitrary spherical shell placed between $r=0$ and $r=r_s$, the pressure is lower than at the center, so that the particles can just be accelerated away from the center. A similar effect of a pressure gradient is present in the Stephani model. Of course, for full analogy one needs expansion scalar $\Theta$ not to vanish, but this surely shows that an inhomogeneous pressure universe can be considered as a kind of interior of an exotic star.

\section{Exact inhomogeneous pressure cosmologies}
\setcounter{equation}{0}
\label{exact}

\begin{figure}[ht]
 %\unitlength1cm
 \begin{center}
 \scalebox{0.4}{\includegraphics[angle=0]{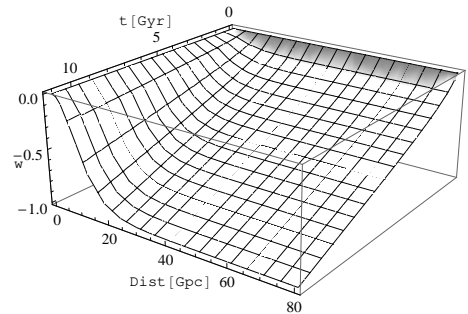}}
\caption{The effective barotropic index (\ref{weff}) as the function of cosmic time in gigayears and the physical distance from the center of symmetry at $r=0$ in gigaparsecs for the spherically symmetric inhomogeneous pressure Stephani model with $\Omega_{inh} = 1 - \Omega_0 = 0.61$ (for this plot we took $w=0$, i.e. the dust matter at the center of symmetry).}
\label{fig1}
 \end{center}
 \end{figure}

In Refs.\cite{dabrowski93,dabrowski95} two exact spherically symmetric Stephani models were found: model I which fulfills the condition $(V/a)^{\cdot \cdot}=0$ and model II which fulfills the condition $(k/a)^{\cdot}=0$ [this reduces the factor in front of $dt^2$ in the metric (\ref{STMET}) just to $-1/V^2$]. A subclass of model II with $k(t) = \beta a(t)$ ($\beta$ = const, with the unit $[\beta]$ = Mpc$^{-1}$) was found in Ref. \cite{stelmach01} and it was assumed that at the center of symmetry the standard barotropic equation of state $p(t) = w \varrho(t)$
%\be
%p(t) = w \varrho(t)
%\ee
was fulfilled. This assumption gives that
\be
\label{SJmod}
\frac{8\pi G}{3} \varrho(t) = \frac{A^2}{a^{3(w+1)}(t)} \hspace{0.3cm} (A = {\rm const.})
\ee
and allows one to write a generalized Friedmann equation as
\be
\label{SJmod2}
 \left(\frac{\dot{a}(t)}{a(t)} \right)^2 = \frac{A^2}{a^{3(w+1)}(t)} - \frac{\beta}{a(t)}
\ee
with the equation of state
\be
\label{weff}
p(t) = \left[w + \frac{\beta}{4} (w+1) a(t) r^2 \right] \varrho(t) = w_{e} \varrho(t)~~.
\ee
Since the effective barotropic index is both timely and spatially dependent here, then it is useful to plot it as the function of these coordinates and it is given in Fig. \ref{fig1}. As we can see the effective barotropic index is getting more and more negative simulating the dark energy for large distances from the center of symmetry and far from the big-bang singularity which is at $t=0$.
Similarly as in the Friedmann models, one can define the critical density as $\varrho_{cr}(t) = (3/8\pi G)[\dot{a}(t)/a(t)]^2~,$
%\be
%\label{rhocrit}
%\varrho_{cr}(t) = \frac{3}{8\pi G} \left(\frac{\dot{a}(t)}{a(t)} \right)^2~,
%\ee
and the density parameter $\Omega(t) = \varrho(t)/\varrho_{cr}(t)$. After taking $t = t_0$, we have from (\ref{SJmod})  that
\be
\label{defOM}
1 = \frac{A^2}{H_0^2 a^{3(w+1)}(t_0)} - \frac{\beta}{H_0^2 a_0} \equiv \Omega_0 + \Omega_{inh}~~,
\ee
and so
\be
\label{beta}
\beta = a_0 H_0^2 \left( \Omega_0 - 1 \right) < 0~~.
\ee

Here we have a two-component universe with the standard matter described by a barotropic equation of state and an inhomogeneity-related kind of exotic matter. In principle, one could consider a multicomponent universe with many different fluids (radiation, stiff matter, etc.) as it was done in an LTB model in Ref. \cite{marra}. However, the point is that models (\ref{SJmod2}) have the property that at the center of symmetry the barotropic equation of state is admitted, and so they include dust (as in LTB models) in a natural way together with an accelerating fluid in one analytic relation. This is a kind of "two-in-one" fluid with dust dominating in one regime and inhomogeneity (or pressure gradient) dominating in the other regime. The same happens in Friedmann-Robertson-Walker models with SFS - in late times some exotic fluid dominates while in early times the standard dust takes over \cite{SFSob}. Also, in a similar way it is possible to simulate dark matter and dark energy with one fluid in $f(R)$ gravity \cite{Noj}.

The nonvanishing components of the 4-velocity and the 4-acceleration vectors are \cite{dabrowski95}
\be
\label{velo}
  u_{\tau}  =  -~\frac{1}{V}~,\hspace{0.8cm} \dot{u}_{r}  =  -~\frac{V_{,r}}{V}~.
\ee
The acceleration scalar is
\begin{equation}
\dot{u}~\equiv \left( \dot{u}_{a} \dot{u}^{a} \right) ^{\frac{1}{2}}~=~
\frac{V_{,r}}{a}~=~~\frac{1}{2} \beta r ,
\end{equation}
and it does not depend on the time coordinate at all. Bearing in mind that the constant $\beta$ is negative in our model [cf. formula (\ref{beta}) for $\Omega_0 < 1$], we have that the highest pressure is at $r=0$ (the center of symmetry), while the lower (negative) pressure regions are outside the center, so that the particles are
accelerated away from the center which is a similar effect as the effect of the positive cosmological constant in the $\Lambda$CDM model. However, in $\Lambda$CDM the pressure is constant everywhere while in the spherically symmetric Stephani model it depends on the radial coordinate $r$. The components of the vector tangent to a null geodesic are \cite{dabrowski95}
\be
\label{COM}
  k^{t}  =  \frac{V^{2}}{a}~, \hspace{0.2cm}
  k^{r} =  \pm \frac{V^{2}}{a^{2}} \sqrt{1~-~\frac{h^{2}}{r^{2}}}~,\hspace{0.2cm}
  k^{\theta}  =  0~,\hspace{0.2cm}
  k^{\phi}  = h \frac{V^{2}}{a^{2}r^{2}}~,
\ee
where $h$ = const., and the plus sign applies to a ray moving away from
the center of symmetry, while the minus sign applies to a ray moving towards the center.
The constant $h$ and the angle $\phi$ between the direction of observation and the direction defined by the
observer and the center of symmetry are related by
\be
\cos{\phi} = \pm \sqrt{1~-~\frac{h^{2}}{r^{2}}}~~.
\ee
The angle $\phi$ should be taken into account when one considers off-center observers \cite{dabrowski95}.

In Ref.\cite{GSS} it was shown that in the model (\ref{SJmod}) the inhomogeneity could mimic the dark energy in the sense that they produce the same redshift-magnitude relation which corresponds to an accelerated expansion of the universe and that $\Omega_{inh,0} = 0.61^{+0.08}_{-0.10}$.
It also emerged that the inhomogeneity had dominated the universe quite recently, so it influenced only slightly the Doppler peaks and did not influence big-bang nucleosynthesis at all. Models of type I have been studied in Ref. \cite{chris}, where they were tested against cosmic microwave background data.

\section{Redshift drift in a pressure-gradient cosmology}
\setcounter{equation}{0}
\label{drift}

Recently, lots of interest was attracted by the effect of redshift drift in cosmological models -- the effect first noticed by Sandage and later explored by Loeb \cite{sandage+loeb}. The idea is to collect data from the two light cones separated by 10-20 years to look for the change in redshift of a source as a function of time. It has recently been investigated for the LTB models \cite{UCETolman}, backreaction timescape cosmology \cite{wiltshire}, and very recently for the axially symmetric Szekeres models \cite{marieN12}. Here we will consider this effect for the Stephani models.

In order to do that we assume that the source does not possess any peculiar velocity, so that it maintains a fixed comoving coordinate $dr=0$. The light emitted by the source at two different times $t_e$ and $t_e+\delta t_e$ will be observed at $t_o$ and $t_o+\delta t_o$ related by
\be
\int_{t_e}^{t_o}\frac{dt}{a(t)}=\int_{t_e+\delta t_e}^{t_o+\delta t_o}\frac{dt}{a(t)}~.
\ee
For small $\delta t_e$ and $\delta t_o$ we have
\be
\label{rel}
\frac{\delta t_e}{a(t_e)}=\frac{\delta t_o}{a(t_o)}~.
\ee
A general formula for redshift which is valid for any cosmological model reads as \cite{KS}
\be
\label{redshift}
1+z = \frac{(u_a k^a)_e}{(u_a k^a)_o}~~.
\ee
The redshift drift is defined as \cite{sandage+loeb}
\begin{eqnarray}
\label{redshiftdrift}
\delta z=\frac{(u_a k^a)(r_e,t_e+\delta t_e)}{(u_a k^a)(r_0,t_0+\delta t_0)}-\frac{(u_a k^a)(r_e,t_e)}{(u_a k^a)(r_0,t_0)}~,
\end{eqnarray}
which can be calculated using the expansions
\bea
\label{taylor1}
(u_a k^a)_o&=& (u_a k^a)(r_0,t_0)+\left[\frac{\partial(u_a k^a)}{\partial t}\right]_{(r_0,t_0)}\delta t_0, \nonumber\\
%(u_a k^a)(r_0,t_0+\delta t_0)
\label{taylor2}
(u_a k^a)_e&=&(u_a k^a)(r_e,t_e)+\left[\frac{\partial(u_a k^a)}{\partial t}\right]_{(r_e,t_e)}\delta t_e~. \nonumber
%(u_a k^a)(r_e,t_e+\delta t_e)
\eea
From (\ref{velo}) and (\ref{COM}) we have
\begin{eqnarray}
\label{rs}
u_a k^a=-\frac{1+\frac{1}{4}k(t)r^2}{a(t)}~.
\end{eqnarray}
Using (\ref{rel}), (\ref{redshift}) and (\ref{rs}), one can calculate the redshift drift (\ref{redshiftdrift}) for the Stephani universes as
\be
\label{redshiftdrift2}
\frac{\delta z}{\delta t}=-\frac{H_0}{1+\frac{1}{4}k(t_0)r_0^2}\left[\frac{H}{H_0} -(1+z) \right]~,
\ee
which with the help of (\ref{defOM}) can be rewritten to the form
\bea
\label{redshiftdrift3}
\frac{\delta z}{\delta t}=-\frac{H_0}{1+\frac{1}{4}H_0^2 (\Omega_0-1)\tilde{r}_0^2} \times \nonumber \\
\left[\sqrt{\Omega_0 \tilde{a}^{-3(w+1)}+(1-\Omega_0) \tilde{a}^{-1}} -(1+z)\right]~,
\eea
where  $\tilde{a}=a/a_0$ and $\tilde{r}=r a_0$.

Using all the above equations (\ref{rel})-(\ref{redshiftdrift3}) we end up with the set of equations which combined together allow us to find the redshift drift of any source at redshift $z$ in the considered class of Stephani models II defined by the condition that $k(t)=\beta a(t)$ as
\bea
\label{redshiftdrift4}
\frac{\delta z}{\delta \tau}&=&-H_0 \frac{4+H_0^2(\Omega_0-1)\tilde{a}\tilde{r}_0^2}{4+H_0^2 (\Omega_0-1)\tilde{r}_0^2} \times \nonumber \\
&& \left(\sqrt{\Omega_0 \tilde{a}^{-3(w+1)}+(1-\Omega_0) \tilde{a}^{-1}} -1-z\right) \\
\label{rs2}
\tilde{a}^{-1}&=&\frac{4+H_0^2(\Omega_0-1)\tilde{r}_0^2}{4+H_0^2(\Omega_0-1)\tilde{a}\tilde{r}^2} (1+z),\\
\label{dtr}
\frac{d\tilde{r}}{d\tau}&=&\pm\left(\tilde{a}^{-1}+\frac{H_0^2}{4}(\Omega_0-1)\tilde{r}_0^2\right)\left(1-\frac{\tilde{r}_0^2}{\tilde{r}^2 } sin^2\phi\right)^{1/2}, \nonumber \\
&&
\eea
where $\tau$ is the proper time ($d\tau = dt/V$) measured by an observer placed at $\tilde{r}_0$ and the last equation describes the propagation of the null geodesic.

In the limit $\Omega_0 \to 1 \Rightarrow \Omega_{inh} \to 0$ and $w=0$ (a flat Friedmann model filled with dust), the formula (\ref{redshiftdrift2}) reduces to
\be
\label{limit1}
\frac{\delta z}{\delta t}=-H_0[(1+z)^{3/2}-(1+z)]~,
\ee
which coincides with the standard Friedmann universe formula obtained by Sandage and Loeb \cite{sandage+loeb}. On the other hand, for an inhomogeneity-dominated universe $\Omega_0 \to 0 \Rightarrow \Omega_{inh} \to 1$, and we have a simple result
\be
\label{limit0}
\frac{\delta z}{\delta t} = H_0 \frac{z}{2}~,
\ee
which means that the drift grows linearly with redshift.
%For the computational convenience we may transform (\ref{redshiftdrift4}), (\ref{rs2}), and (\ref{dtr}) to
%\begin{eqnarray}
%\label{redshiftdrift5}
%\frac{\delta z}{\delta \tau}&=&-\frac{H_0}{1+\frac{1}{4}H_0^2 (\Omega_0-1)\tilde{r}_0^2} \times \\
%&& \left[\sqrt{\Omega_0 \tilde{a}^{-3(w+1)}+(1-\Omega_0) \tilde{a}^{-1}} -(1+z) \right] \nonumber \\
%\label{rs3}
%\tilde{a}^{-1}&=&\left[1+\frac{H_0^2}{4}(\Omega_0-1)\tilde{r}_0^2\right](1+z)-\frac{H_0^2}{4}(\Omega_0-1)\tilde{r}^2,\\
%\label{dtr1}
%\frac{d\tilde{r}}{dz}&=&\frac{1+\frac{H_0^2}{4}(\Omega_0-1)\tilde{r}_0^2}{H_0\sqrt{\frac{\Omega_0 \tilde{a}^{-3(w+1)}+(1-\Omega_0) %\tilde{a}^{-1}}{1-\frac{\tilde{r}_0^2}{\tilde{r}^2}sin^2\phi}}+\frac{H_0^2}{2}(\Omega_0-1)\tilde{r}}~~.
%\end{eqnarray}
%
%
\begin{figure}[ht]
 %\unitlength1cm
 \begin{center}
 %\scalebox{1.5}{\includegraphics[width=7cm,height=5.5cm,angle=270]{sfstoz10.eps}}
 \scalebox{1.0}{\includegraphics[angle=0]{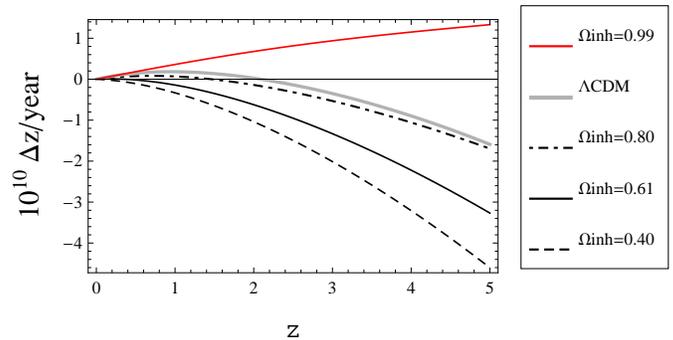}}
\caption{The redshift drift (\ref{redshiftdrift3}) as a function of redshift for the sperically symmetric inhomogeneous pressure Stephani model with $r_0=0$, $w=0$, and $\Omega_{inh}=1-\Omega_0=0.40;0.61;0.80;0.99$. It is clear that the drift is similar to the (negative) drift of LTB models if the parameter of inhomogeneity $\Omega_{inh}$ is small, while it is like the drift in $\Lambda$CDM models for larger values of the inhomogeneity parameter $\Omega_{inh}$. For larger values of redshift both Stephani and $\Lambda$CDM models behave similar to the void LTB models and the drift becomes negative. However, for a very large inhomogeneity, the  drift becomes positive for larger and larger redshifts, reaching the limit that the inhomogeneity is totally dominating $\Omega_{inh} = 1$ with linear drift dependence on $z$ [cf. Eq. (\ref{limit0})] being always positive. The plot shows that one is able to differentiate between the drift in $\Lambda$CDM models and in Stephani models which can be verified in future experiments \cite{balbi}. It also shows that the LTB inhomogeneity (due to the energy density) is different from the Stephani inhomogeneity (due to the pressure) which exhibits as the fact that the drift is always negative for an LTB model and always positive for an inhomogeneity-dominated Stephani model.}
\label{fig2}
 \end{center}
 \end{figure}

 \begin{figure}[ht]
 %\unitlength1cm
 \begin{center}
 %\scalebox{1.5}{\includegraphics[width=7cm,height=5.5cm,angle=270]{sfstoz10.eps}}
 \scalebox{1.0}{\includegraphics[angle=0]{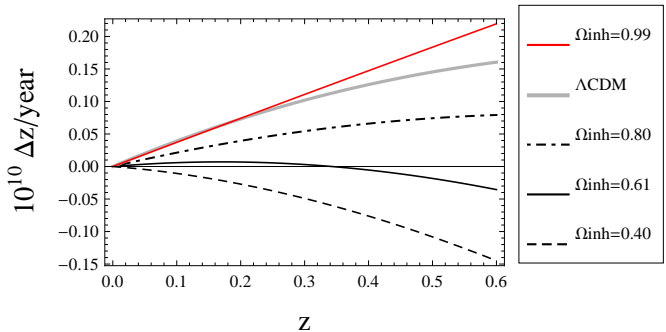}}
\caption{The same plot of redshift drift for the spherically symmetric inhomogeneous pressure Stephani model with $r_0=0$, $w=0$, and $\Omega_{inh}=1-\Omega_0=0.40;0.61;0.80;0.99$ as in (\ref{fig1}), but enlarged for small redshifts. It is seen that small inhomogeneity (e.g. $\Omega_{inh}=0.40$) makes the model to behave almost like LTB, while large inhomogeneity makes it more similar to $\Lambda$CDM.}
\label{fig3}
 \end{center}
 \end{figure}

%\begin{figure}[ht]
 %\unitlength1cm
% \begin{center}
 %\scalebox{1.5}{\includegraphics[width=7cm,height=5.5cm,angle=270]{sfstoz10.eps}}
% \scalebox{0.6}{\includegraphics[angle=0]{delta2.eps}}
%\caption{The plot of redshift drift (\ref{redshiftdrift3}) as a function of redshift for the sperically symmetric inhomogeneous pressure Stephani model with $r_0=0$, %$w=0$, and $\Omega_{inh}=0.61$ only. The drift is positive for the redshifts $z\in(0, 0.34)$, and reaches its highest value $\approx 10^{-12}/year$ at $z\sim 0.17$.}
%\label{fig3}
% \end{center}
% \end{figure}

In Ref.\cite{Quercellini12} the redshift drift as a function of redshift for the $\Lambda$CDM model, the Dvali-Gabadadze-Porrati (DGP) brane model, the matter-dominated model (CDM), and other three different LTB void models were presented. It has been shown that the redshift drift $\delta z / \delta t$ for $\Lambda$CDM and DGP models is positive up to $z \approx 2$ and becomes negative for larger redshifts, while it is always negative for LTB void models \cite{yoo}. Using the formulas ({\ref{redshiftdrift4})-(\ref{dtr}) we plot the redshift drift as a function of redshift for the Stephani model with $r_0=0$, $w=0$, and $\Omega_{inh}=1-\Omega_0=0.40;0.61;0.80;0.99$ in Figs. \ref{fig2} and \ref{fig3}. It emerges that for large redshifts the drift for Stephani models and $\Lambda$CDM models exhibits the behavior which is like the redshift drift in the void LTB models. However, unlike in the void models, and depending on the value of inhomogeneity $\Omega_{inh}$, it becomes positive for small redshifts and approaches the behavior of the $\Lambda$CDM model, which allows negative values of the drift, for very high redshifts. For example, in the model with $\Omega_{inh}=0.61$ the redshift drift becomes positive for $z\in(0, 0.34)$ and attains its highest value $ \approx$ 1.0 $\cdot$ 10$^{-12}$/year for $z\sim 0.17$ (cf. Fig. \ref{fig3}). However, for a very large inhomogeneity, the  drift becomes positive for larger and larger redshifts (e.g. for $\Omega_{inh} = 0.99$ it is positive until $z \approx 13$), reaching the limit that the inhomogeneity is totally dominating $\Omega_{inh} = 1$ with linear drift dependence on $z$ [cf. Eq. (\ref{limit0})] being always positive. This may allow one to detect the difference between the spherically symmetric LTB models and spherically symmetric Stephani models as well as the $\Lambda$CDM models in future telescopes such as the European Extremely Large Telescope (EELT) (with its spectrograph CODEX (COsmic Dynamics EXperiment)) \cite{balbi,E-ELT}, the Thirty Meter Telescope (TMT), the Giant Magellan Telescope (GMT), and especially, in gravitational wave interferometers DECIGO/BBO (DECi-hertz Interferometer Gravitational Wave Observatory/Big Bang Observer) \cite{DECIGO}. The first class of the experiments involving the very sensitive spectrographic techniques such as those utilized in the CODEX spectrograph use a detection of a very slow time variation of the Lyman-$\alpha$ forest of the number of quasars uniformly distributed all over the sky to measure the redshift drift.  However, since Lyman-$\alpha$ lines become impossible to measure for  $z<1.7$ from the ground \cite{E-ELT}, such experiments are incapable to distinguish between the void LTB models and the mimicking dark energy Stephani models with $\Omega_{inh}=0.61$. On the other hand, such a distinction seems to be possible to make with the other mentioned class of future experiments involving the space-borne gravitational wave interferometers DECIGO/BBO. Such experiments are based on the measurement of the correction due to the accelerating expansion of the Universe to the phase of the hypothetical gravitational waves coming from neutron-star binaries. As was shown in  Ref. \cite{DECIGO}, a detection of such a phase correction  may be used to infer the positivity of the redshift drift at even $z\sim 0.2$. This suggests that with the future observations of gravitational waves it will be possible to rule out any void LTB models, unless one assumes an unrealistically steep density gradient for $z\sim 0$. In this regard, the future experiments  involving  the gravitational wave interferometers DECIGO/BBO may be thought to be complementary to all those experiments that use the shift of the Lyman-$\alpha$ forest to detect the redshift drift. Besides, it is worth mentioning that there is another test known as the cosmic parallax test \cite{Quercellini12} which is due to anisotropic expansion and is strictly related to a nonvanishing shear. However, Stephani universes as shear-free should not experience it - this may be another way to differentiate Stephani models and LTB models.

\section{Conclusions}
\setcounter{equation}{0}
\label{conclusion}

Observations from just point in the Universe we make suggest its isotropy, but not necessarily homogeneity. This gives some motivation for studying inhomogeneous spherically symmetric models of the Universe rather than isotropic and homogeneous ones. In this paper we have discussed the Stephani models of pressure-gradient spherical shells which are complementary to the energy density varying spherical shells of the Lema\^itre-Tolman-Bondi models. The formula for redshift drift $\delta z/ \delta t$ of any source at redshift $z$ in the specific class of Stephani models for both centrally and noncentrally placed observers has been obtained. We have shown that, at least for the centrally placed observers, there is a subset of observationally viable Stephani models which exhibit qualitatively different behavior of redshift drift than the LTB void models as well as the quantitatively different behavior than the $\Lambda$CDM models. We proved that small inhomogeneity (e.g. $\Omega_{inh}=0.40$) makes a Stephani model behave almost like an LTB model, while large inhomogeneity makes it behave more similar to $\Lambda$CDM. For a very large inhomogeneity, the  drift becomes positive for larger and larger redshifts, reaching the limit that the inhomogeneity is totally dominating $\Omega_{inh} = 1$ with linear drift dependence on $z$ [cf. Eq. (\ref{limit0})] being always positive. This gives a good perspective to differentiate between the drift in $\Lambda$CDM models and in Stephani models in future experiments \cite{balbi}. It is vital that the LTB inhomogeneity (due to the energy density) is different from the Stephani inhomogeneity (due to the pressure) which exhibits as the fact that the drift is always negative for an LTB model and always positive for an inhomogeneity-dominated Stephani model. These differences may allow one to test Stephani cosmology against LTB and $\Lambda$CDM cosmology in future experiments aimed to measure the redshift drift, especially those aimed at low redshift like DECIGO/BBO.

If the observations show that the drift is positive, then the data will exclude the void LTB models which allow the negative redshift drift only, while this will not be the case for both $\Lambda$CDM and Stephani models. The difference between $\Lambda$CDM and Stephani models will be determined, provided the inhomogeneity is not large. On the other hand, if the data show that the drift is negative, then both $\Lambda$CDM and Stephani models will have to be rejected. This gives a clear test for all three models of the Universe.

\section{Acknowledgements}

The authors acknowledge the discussions with Marie-No\"elle C\'el\'erier and David Polarski. This work was supported by the National Science Center Grant No. N N202 3269 40 (years 2011-2013).

\end{document}